\documentclass[sigconf]{acmart}

\AtBeginDocument{%
  \providecommand\BibTeX{{%
    \normalfont B\kern-0.5em{\scshape i\kern-0.25em b}\kern-0.8em\TeX}}}

\setcopyright{acmcopyright}
\copyrightyear{2018}
\acmYear{2018}
\acmDOI{10.1145/1122445.1122456}

\acmConference[Woodstock '18]{Woodstock '18: ACM Symposium on Neural
  Gaze Detection}{June 03--05, 2018}{Woodstock, NY}
\acmBooktitle{Woodstock '18: ACM Symposium on Neural Gaze Detection,
  June 03--05, 2018, Woodstock, NY}
\acmPrice{15.00}
\acmISBN{978-1-4503-XXXX-X/18/06}

\usepackage{graphicx}
\usepackage{amsmath}
\usepackage{amsthm}
\usepackage{booktabs}
\usepackage{adjustbox}
\usepackage{subcaption}
\usepackage{float}
\usepackage{color}
\usepackage{amsmath}

\usepackage{balance}
\title{SCNet: A Neural Network for Automated Side-Channel Attack}

\author{Guanlin Li}
\affiliation{
\institution{
Shandong Computer Science Center}
}
\email{leegl@sdas.org}

\author{Chang Liu}
\affiliation{%
  \institution{Nanyang Technological University}
  }
\email{chang015@e.ntu.edu.sg}

\author{Han Yu}
\affiliation{%
  \institution{Nanyang Technological University}
  }
\email{han.yu@ntu.edu.sg}

\author{Yanhong Fan}
\affiliation{%
  \institution{Shandong University}
}
\email{fanyh@sdas.org}

\author{Libang Zhang}
\affiliation{%
 \institution{Tsinghua University}
}
\email{zlb17@mails.tsinghua.edu.cn}

\author{Zongyue Wang}
\affiliation{%
  \institution{Open Security Research Company}
}
\email{zongyue.wang@opsefy.com}

\author{Meiqin Wang}
\affiliation{%
  \institution{Shandong University}
  }
\email{mqwang@sdu.edu.cn}

\begin{document}

\begin{abstract}
The side-channel attack is an attack method based on the information gained about implementations of computer systems, rather than weaknesses in algorithms. Information about system characteristics such as power consumption, electromagnetic leaks and sound can be exploited by the side-channel attack to compromise the system. Much research effort has been directed towards this field. However, such an attack still requires strong skills, thus can only be performed effectively by experts. Here, we propose SCNet, which automatically performs side-channel attacks. And we also design this network combining with side-channel domain knowledge and different deep learning model to improve the performance and better to explain the result. The results show that our model achieves good performance with fewer parameters. The proposed model is a useful tool for automatically testing the robustness of computer systems. 
\end{abstract}




\maketitle

\section{Introduction}

Deep learning models have been used in many areas, such as image classification~\cite{he_deep_2016,szegedy_inception-v4_2017}, object detection~\cite{lin_feature_2017}, and natural language processing (NLP)~\cite{vaswani_attention_2017,gehring_convolutional_2017}. While deep learning has made huge improvements on computer vision (CV) and NLP, using deep learning to perform security-related tasks is also catching people's attention. Recently, researchers pay more attention to performing cryptanalysis using neural networks, especially side-channel attacks~\cite{maghrebi_breaking_2016,picek_side-channel_2017,wei_i_2018}. We notice that Maghrebi et. al~\cite{maghrebi_breaking_2016} have already compared a lot of machine learning models, such as random forest, support vector machine, convolutional neural network and recurrent neural network. But they evaluate them on DPA\_v2~\cite{danger_education_2011} with very simple structure, which is hard to realize the full potential of neural networks. More than that, they do not design a model for side-channel attack specially.%should have a cite here

The side-channel attack (SCA) based on power consumption~\cite{kocher_differential_1999,chari_template_2002,mangard_simple_2002} is a powerful attack method. It takes advantage of information obtained by the implementation of the security algorithms (i.e. varying power consumption when the circuit runs different operations) in order to obtain part of the secret information. One of the common targets of SCA is the key used in a block cipher~\cite{katz_introduction_2007}. A block cipher consists of an encryption algorithm $E(x,k)$ and a decryption algorithm $D(y,k)$, where $x$ is the plaintext to be encrypted, $k$ is the key, and $y$ is the ciphertext. They are all binary strings with fixed lengths. Generally $x=D(y,k)$ and $y=E(x,k)$. For example in AES-128~\cite{katz_introduction_2007}, $x,y,k$ have same length 128 bits, i.e. 16 bytes.

To obtain the key $k$ fixed in a block cipher running on a cipher chip, here we discuss two main \textbf{threat models} in side-channel attacks based on the power of the attackers: a) The attackers can encrypt any messages they want by any key to sample the leaked information (i.e. power consumption) from the attacked chip. So they can analyze power traces by building a power consumption template~\cite{chari_template_2002,standaert_using_2008} on this chip; and b) The attackers can only encrypt messages by specific key fixed in the attacked chip. So they need to use a same type of chip as a template one to encrypt any messages they want by any key to obtain the leaked information. And they build a consumption model based on known information from the template chip to predict the key fixed in the attacked chip. All these two threat models, the attackers build a reference model based on data from different sources. %Or they can directly use differential power analysis (DPA) or high-order differential power analysis (HO-DPA)~\cite{kocher_differential_1999,kocher_introduction_2011} to attack this chip.}

%These traditional attack methods are able to uncover sensitive information related to the secret key. %However, to attack the block cipher implementation by neural networks, we need to build a model to predict the key by analyzing the correlation between sampling points and the intermediate results (e.g., the power consumption during the encryption process). In this situation, relying on rules to compare attack traces with profiling traces is not sufficient. 

\begin{figure}[ht]
\centering
\includegraphics[width=1.0\columnwidth,height=120 pt]{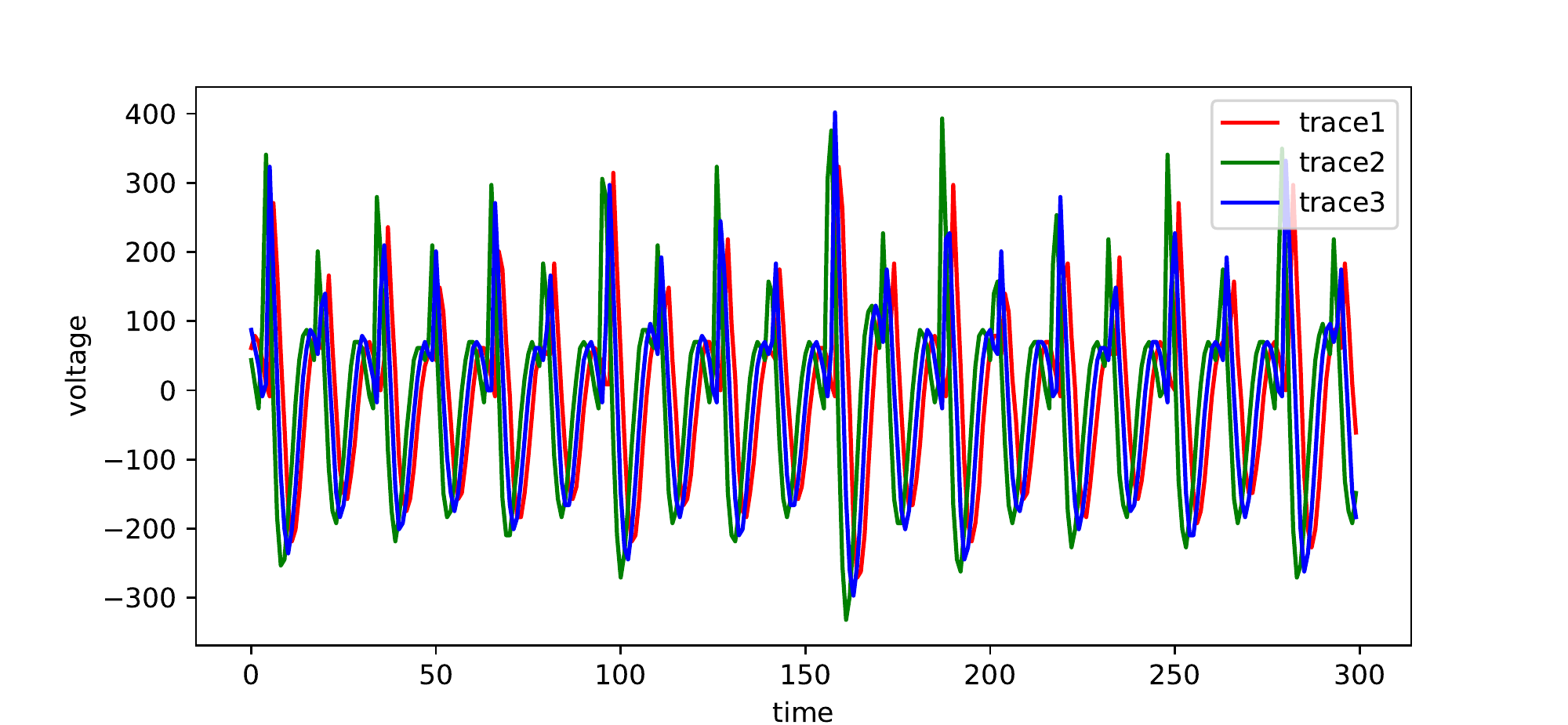}
\caption{The traces are the voltage fluctuations of the cipher chips during the encryption. The side-channel attack models take these as input, and recovers the keys without the intermediate of domain expert.}
\label{fig:traces}
\end{figure}

In this paper, we propose a novel approach under threat model a) to analyze power consumption information (i.e. voltages shown in \textbf{Figure~\ref{fig:traces}}) to attack block ciphers. Here, \textbf{sampling points} refer to the voltages sampled from the cipher chip at any given time during the encryption. The sequence of sampling points is called the \textbf{trace}. %, which contains the information about the operations and operating bits of the encryption process. %However, traces obtained from block ciphers which use different keys often do not exhibit a common pattern. 
\textbf{Figure~\ref{fig:traces}} gives an example of traces which encrypt the same plaintext with different keys. %In side-channel waveforms,
Usually, many sampling points describe one same operation used in block ciphers together. We refer to leakage hidden in these points the \textbf{\textit{crossing information}}, which is similar to the result of the high order differential power attack(HO-DPA) function~\cite{kocher_differential_1999,kocher_introduction_2011} which combines multiple sampling points from one trace. The HO-DPA function can be explained as a function combining multiple points using operations like adding, subtracting and multiplying. But crossing information is made by multiplying self-adaptation weights to a group of sampling points and then applying an HO-DPA function on them. It is difficult to distinguish which group of sample points contains the leaked information of the secret key when the block cipher equips with a defense strategy. In order to solve this problem, we propose the \textbf{sampling point embedding} technique, which can automatically transfer each sampling point into a vector having a fixed length to encode the leaked information behind the point. %Each sampling point embedding vector can be regarded as the output of a noisy word embedding matrix~\cite{mikolov_efficient_2013}, which transforms different operations from codes to vectors. In this way, the sampling points can be regarded as the sum of the elements of corresponding vectors. 
We further propose a new deep learning model, called SCNet, to automatically perform side-channel attacks with good performance. {In SCNet, we propose a dilated encoder block to generate embedding vectors for sampling points. It can obtain much more diverse features and avoid over-fitting on some sensitive noise points. At the end of the block, we stack features from each encoder in this block as a whole one. The model can learn how to generate correct embedding vectors in an end-to-end fashion. }

%It firstly follows the conventional method to denoise the sampling points and find the most related points to predict the block cipher key. %SCNet leverages the Inception~\cite{szegedy_going_2015} structure to reduce the number of model parameters. Its structure is based on ResNeXt~\cite{xie_aggregated_2017}. 

\begin{figure}[ht]
\centering
\includegraphics[width=1.0\columnwidth,height=120 pt]{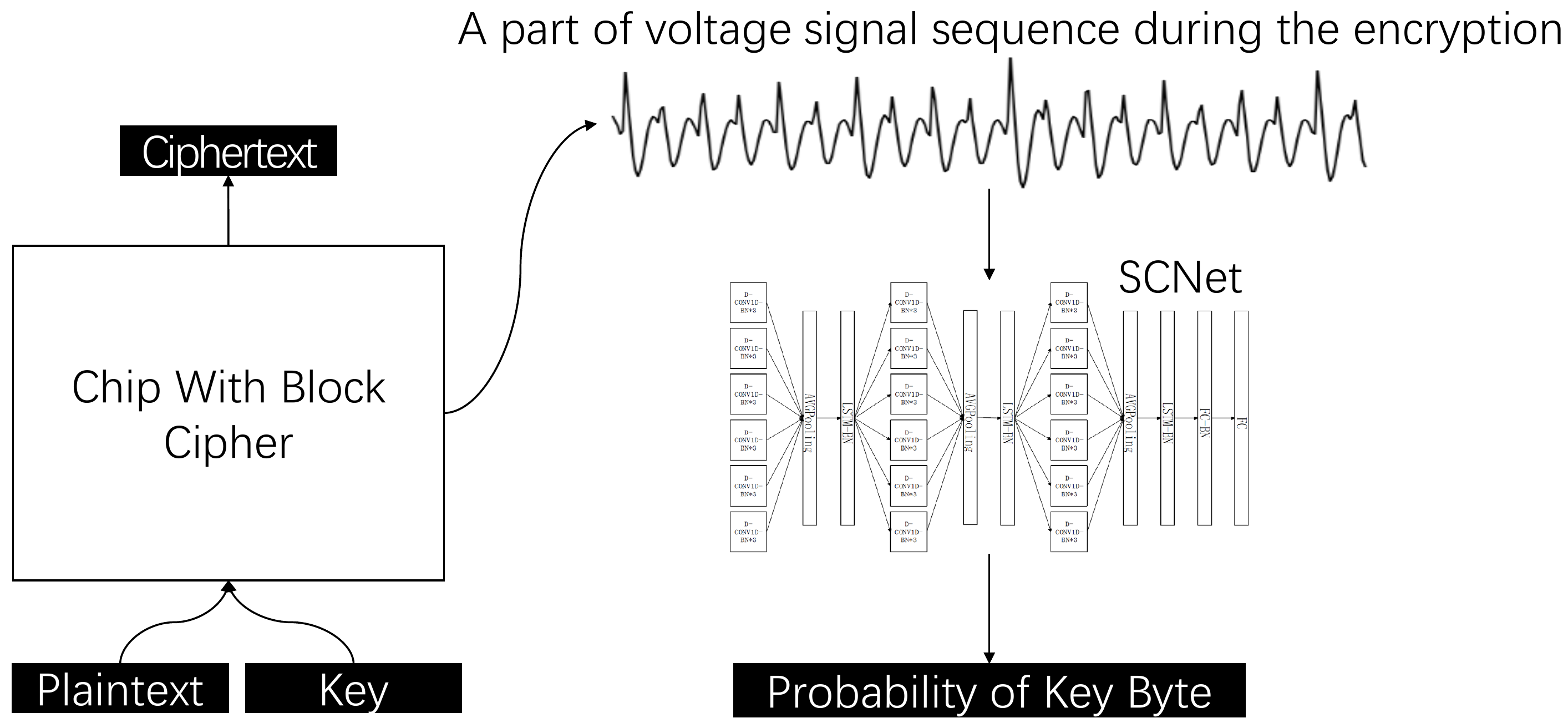}
\caption{An overview of automated side-channel attack by using SCNet. Only the points of interest are kept in each voltage signal sequence (i.e. trace).}
\label{fig:oa}
\end{figure}

In \textbf{Figure~\ref{fig:oa}}, we show a system architecture of predicting the secret key byte by using SCNet. A block cipher algorithm is running in a chip which attackers can input keys and plaintexts and receive ciphertexts, i.e. threat model a) we have introduced. Attackers sample the voltage signals by using a resistor when encrypting messages and choose the sampling points related with the attacked key byte to input to the model, which are called points of interest. Then, the model automatically transfer each sampling point into embedding vector and calculate crossing information which is used to predict the probability of key byte.

The commonly used evaluation method of SCA is based on the success rate (SR) and guessing entropy (GE)~\cite{standaert_unified_2009}, representing whether the model predicts the right label and the average remaining workload to predict the right one, respectively. The idea is to find the SR and GE expectations under multiple power traces through multiple attack experiments. In our experiments, we use this metrics to evaluate our model and others. Mostly, we compare SCNet with ASCAD CNN model~\cite{prouff_study_2018}. According to the result shown in~\cite{prouff_study_2018}, ASCAD CNN has better performance than template attack, VGG-16 and MLP. In addition, we compare SCNet with SCNet{\_seq}, which employs all the components in SCNet but is built in a sequential manner, i.e. without dilated blocks. In practice, we also need to consider the time complexity and storage space complexity of the attack method. Approaches that require less training and predicting data are also considered superior. Thus, we evaluate the models in terms of time consumption, model size, and number of traces required for prediction. More than that, we prefer to introduce a new view to explain how neural network can help us solve this question.

In summary, our contributions in this paper are as follows:
\begin{enumerate}
    \item We propose SCNet. To the best of our knowledge, it is the first time to explain the reason why we can use deep learning models to help us analyze side-channel traces. And it is designed for side-channel attacks specially, so it is more powerful to attack block ciphers with defense.
    \item We design dilated encoder blocks and sampling point embedding techniques to accurately obtain the crossing information between sampling points and the secret key.
    \item Extensive experimental evaluations using real-world public datasets including ASCAD~\cite{prouff_study_2018} and DPA{\_v4.2}~\cite{bhasin_analysis_2014} demonstrate that our approach is reasonable for building the attacking model.
\end{enumerate}

The rest of this paper are organized as follows. Section 2 reviews related works in areas concerning techniques involved in the proposed approach. Section 3 introduces how we assume the properties of the sampling points and propose a new method to analyze the points based on their properties. Section 4 discusses the details of the proposed SCNet. Section 5 introduces the experiment settings and interprets the experimental evaluation results. Finally, in Section 6, we conclude the paper and discuss potential future research directions. And you can find our models and dataset from https://github.com/GuanlinLee/SCNet.

\section{Related Work}
In this section, we review related work in areas concerning techniques involved in the proposed approach.

\subsection{Side-Channel Attacking Technologies}

\subsubsection{Differential Power Analysis (DPA)}
DPA measures power levels at different parts of the cipher chip and applies statistical analysis to overcome countermeasures, such as added noise, that are applied to obscure individual bits~\cite{kocher_differential_1999,kocher_introduction_2011}. Specific operating information can be obtained by DPA to recover the secret key. Firstly, the attacker obtains many encryption traces about different plaintexts at random. Some of the traces are related with one specific bit, which is label 0; while others are related with label 1 of this bit. The positions having huge difference between 0's traces and 1's traces are the most related to the secret information. Then, the attacker can guess some values of the secret key and find the most related one.
%这段的描述需要修改，reviewer认为是confusing的

\subsubsection{Correlation Power Analysis (CPA)}
CPA was proposed by~\cite{brier_correlation_2004}. To predict the secret key, the adversary needs to model the leaking information. Usually, the adversary analyzes the correlation between a distribution $t$ of sampling points and a distribution $HW(y)$ of the Hamming Weight of the intermediate results of cryptography by using:
\[C(t,y)=\frac{E[t*HW(y)]-E[t]E[HW(y)]}{\sqrt{Var[t]Var[HW(y)]}}.\]
By analyzing $C(t,y)$, the adversary can find the most related intermediate results to obtain the secret key.
Firstly, we need to measure the actual power consumption of the chip when it encrypts multiple different plaintexts. Then, we calculate  power consumption which we guess is true according to the power leakage model~\cite{messerges_investigations_1999}. Finally, in order to restore the secret key, we analyze the correlation between the two kinds of power consumption.

%For analyzing the correlation, the Pearson product-moment correlation coefficient is required:
%\[C(T,P)=\frac{E[TP]-E[T]E[P]}{\sqrt{Var[T]Var[P]}}.\]
%Here, $T$ represents the set of the actual power consumption corresponding to the key. $P$ represents the set of the hypothetical power consumption corresponding to the guessed key. $E[x]$ represents the expectation of $x$, and $Var[x]$ represents the variance of $x$. If $C(T, P) = 0$ is true, there is no correlation between $T$ and $P$. Otherwise, there is correlation between $T$ and $P$, and the correctly guessed key corresponds to the maximum absolute value of $C(T, P)$.

\subsubsection{Template Attack}
Template attack is a powerful type of side-channel attack. In order to implement a template attack, the attacker first needs to create templates of different secret keys~\cite{standaert_using_2008} by finding a set of functions to fit the collected traces with different keys. It is common for an attacker to analyze a same device as the one being attacked. By modeling and analyzing the templates, the traces of the attacked device are further compared, and the key information can be obtained by using maximum likelihood analysis.

\subsubsection{ASCAD}
Prouff et al.~\cite{prouff_study_2018} proposed the ASCAD dataset, which is a public side-channel attack dataset. In this dataset, SCA is implemented by the Advanced Encryption Standard (AES)~\cite{katz_introduction_2007}. A mask is employed to lower the correlation between the power consumption and the intermediate value of the algorithm by randomizing certain values in the AES program. The traces of masked AES in the dataset are synchronized, and no specific hardware countermeasure is activated on the ATMega8515. Only the 700 points of interest are kept in each trace. There are 50,000 traces in the train set and 10,000 traces in the testing set. They propose to use AlexNet to perform key recovery with the 2-D convolution replaced by a 1-D convolution. They compared their model with a baseline neural network (e.g. vanilla NN) and conventional methods. The results show that CNNs are much better than NNs and conventional methods.

\subsection{Deep Learning Technologies}

\subsubsection{Neural Network}
Neural network(NN) is models composed of neural units. Units are separated into many groups. Each group corresponds to a layer. NN may have hundreds of layers in a model, which includes millions of units in it. After each layer there is an non-linear activation function, e.g., Sigmoid, Tanh, ReLU~\cite{dahl_improving_2013}, which introduce more non-linear characteristic to the model. 

\begin{align*}
    &\mathrm{Sigmoid}(x)=\frac{1}{1+\exp{(-x)}}\\
    &\mathrm{Tanh(x)}=\frac{\exp{(x)}-\exp{(-x)}}{\exp{(x)}+\exp{(-x)}}\\
    &\mathrm{ReLU(x)}=max(x,0)
\end{align*}

To train a NN having ability to achieve special task, we need a dataset named train set. And to validate the generalization of other datasets, we also need a valid set and a test set. During the training process, we feed the network with input data from train set to calculate the output. And the label from train set and the output are sent to the loss function which is used to obtain a objective value. Then a process named backward propagation~\cite{lecun_efficient_2012} starts to compute the gradients of the units layer by layer and use the gradients to update each unit's value. After training, the network would have a good performance on both valid set and test set.

\subsubsection{Dilated Convolutions}
The main application of the dilated convolution is the density prediction: computer vision applications where the projected object has a similar size and structure to the input image~\cite{wang_understanding_2018}. The filters of dilated convolution are similar to the ones in ordinary convolution, while they skip some points during convolution, which directly lead to the increase of the receptive field. In implementations, it can also decrease the number of parameters ensuring that the model has much lower over-fitting risk. By using dilated convolution, the convolutional layer can capture farther spatial information and richer semantic information.

\subsubsection{Long Short Term Memory (LSTM)}
LSTM is a kind of the recurrent neural network (RNN), which is usually used to handle sequential data. A standard LSTM unit is called a cell, which has an input gate, an output gate, and a forget gate~\cite{hochreiter_long_1997}. The cell remembers values over arbitrary time intervals and the three gates regulate the flow of data into and out of the cell, which can reduce the influence of gradient vanishing. LSTM is currently widely used in text translation, speech synthesis, and other fields. With the deepening of the research, there are various versions of LSTM, such as GRU~\cite{chung_empirical_2014} and Bidirectional LSTM~\cite{graves_framewise_2005}.

%\subsubsection{Word Embedding}
%Word Embedding takes a large corpus of text as its input, and produces a vector space, typically of several hundred dimensions, with each unique word in the corpus being assigned a corresponding vector in the space~\cite{goldberg_word2vec_2014}. Word vectors are positioned in the vector space such that words that share common contexts in the corpus are located in close proximity to one another in the space. By computing the vector representation of the word, semantic information and syntax can be reproduced.

%To the best of our knowledge, the SCNet proposed in this paper is the first deep learning-based approach to effectively automate side-channel attacks.

\section{Attacking assumptions}
In this section, we introduce some assumptions of the traces, i.e. sampling point sequences. In~\cite{kocher_differential_1999}, a DPA test can be summarized as follows: 
let $T_{[1,\dots,m]}[1,\dots,k]$ denote $m$ traces, which consist of $k$ sampling points. Let $T_{c}[i]$ denote the $i$-th point within the trace $T_c$. Let $C_{[1,\dots,m]}$ denote $m$ known inputs or outputs for the traces with $C_c$ corresponding to $T_c$. Let $B(C_c, K_n)$ denote a binary valued selection function with input $C_c$ and $K_n$ as the guessed key byte. Each point $i$ in the differential trace $\Delta_{B_i}$ for the guess $K_n$ is computed as follows:
\[\Delta_{B_i}=\frac{\sum_{c=1}^{m}B(C_{c},K_{n})T_{c}[i]}{\sum_{c=1}^{m}B(C_{c},K_{n})}-\frac{\sum_{c=1}^{m}(1-B(C_{c},K_{n}))T_{c}[i]}{\sum_{c=1}^{m}(1-B(C_{c},K_{n}))}.\]
We use the traditional methods to guide us to analyze the traces, but using a neural network.

\subsection{Waveform Resolving}
Suppose that the traces comprise of various kinds of operating information, such as XOR and S-Box( i.e. a nonlinear transform)~\cite{katz_introduction_2007}, which satisfies:
\begin{align*}
    x_{i}&=\sum^{p-1}_{j=0}\sum^{N-1}_{l=0}x_{i,j,l}+\sum^{p-1}_{l=0}\varepsilon_{i,l} ,\\
    x_{i,j}&=[x_{i,j,0},x_{i,j,1},\dots,x_{i,j,N-1}], \\
    \varepsilon_i&=[\varepsilon_{i,0},\varepsilon_{i,1},\dots,\varepsilon_{i,p-1}],
\end{align*}
where $x_{i,j}$ represents the $j$-th operating information concerning the $i$-th sampling point, and $x_{i,j}$ is an $N$-dimensional vector. $p$ is the number of different operations. $\varepsilon_i$ is an $p$-dimensional noise vector. For every sampling point, there is a noise feature for each operation vector. $x_i$ is a direct representation for the information carried by the $i$-th sampling point.

For the aligned sub-traces $X,\ X'$ from two different traces, the sampling points at the same coordinate are related to the encrypting bits. That is to say when the operating bits, i.e. the intermediate results in the block cipher, are the same, the difference between the two sub-traces should only be the noise distribution:
\begin{align*}
X{[a,b]}=[x_a,x_{a+1},x_{a+2},\dots ,x_{b}] , \nonumber\\
X'{[a,b]}=[x'_a,x'_{a+1},x'_{a+2},\dots ,x'_{b}] , \nonumber
\end{align*}
\begin{center}
$\forall \  (x_i - \varepsilon_i ),\  (x'_i - \varepsilon'_i ) \in opset_{i,k},\ \mathrm{for\  the\  same\ intermediate\ results \ in \ the \ block \ cipher},\ x_i-x'_i \sim (\varepsilon_i - \varepsilon'_i) $,
\end{center}
where ${opset}_{i,k}$ denotes a set of the $k$-th$\  (k \le p)$ available operation at position $i$. %We design new network models, which perform waveform resolving and denoising using the signal vector which is similar to the word vector~\cite{goldberg_word2vec_2014}. 
It is assumed that every sampling point can be represented by an $N$-dimensional vector.% To predict the secret key, we need to obtain the clean $N$-dimensional signal vector.

\subsection{Waveform noise model}
\label{sec:2.2}
In template attacks~\cite{standaert_using_2008}, the multivariate Gaussian distribution is proposed to model the noise. However, here we choose to use the Wiener process to model noise to leverage temporal information. For each dimension in the $p$-dimensional vector, we assume that the noise is sampled from a same Wiener process. The Wiener process in each dimension is independent and identical:
\begin{align*}
\centering
    {\varepsilon }_{i,h}\left(t\right)&-{\varepsilon }_{i,h}\left(s\right)\sim N\left(0,{\sigma_h }^{2\left(t-s\right)}\right),\\
&\mathrm{for} \ t>s>0, \ h \le p-1.
\end{align*}

More than that, the noise distribution between sampling points from any two different sub-traces is:
\begin{align*}
\centering
 &\mathrm{\forall }\ (x_i - \varepsilon_i ),\  (x'_i - \varepsilon'_i ) \ \in {opset}_{i,k},\ 
 \mathrm{for \ the \  same }\\ &\mathrm{ intermediate\ results \ in \ the \ block \ cipher,}\ x_i-x'_i \\
 &\sim  N ({\mu }_0,{\mu }_1,{\mu }_2,\dots ,{\mu }_{p-1};{\sigma }_{0}^{2},{\sigma }_{1}^{2},{\sigma }_{2}^{2},\dots ,{\sigma }_{p-1}^{2} ).
\end{align*}

\begin{figure*}[ht]
\centering
\includegraphics[width=1.0\textwidth]{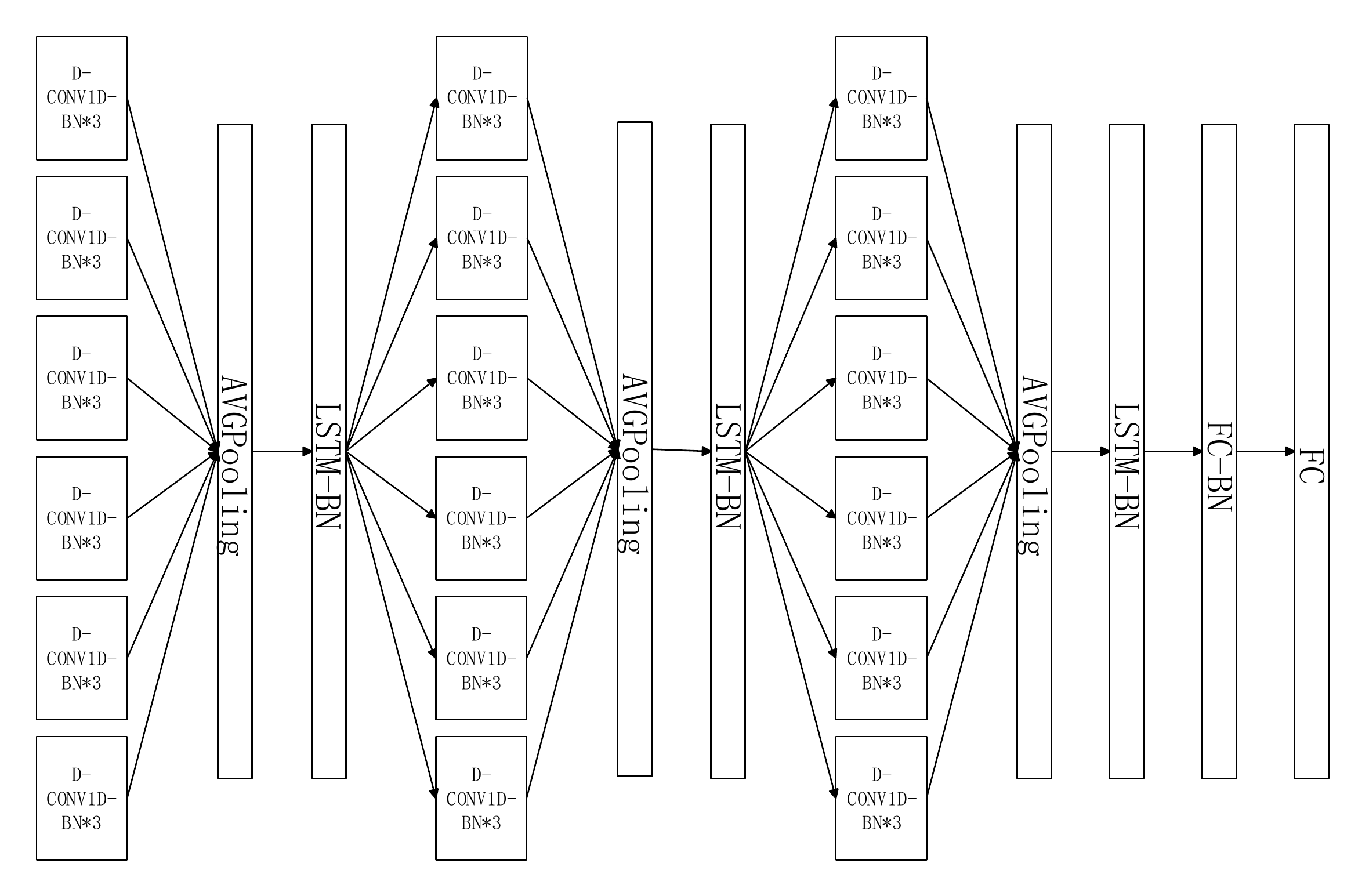}
\caption{The structure of SCNet. \textbf{D-CONV1D} is the one-dimensional dilated convolutional layer. \textbf{BN} is the batch norm layer~\protect\cite{ioffe_batch_2015}. \textbf{AVGPooling} is the average pooling layer. \textbf{LSTM} is the LSTM layer. \textbf{FC} is the fully connected layer.}
\label{fig:2}
\end{figure*}

%\begin{figure*}[ht]
%\centering
%\includegraphics[width=1.0\textwidth,height=0.25\textheight]{image1}
%\caption{The structure of SCNet\_{seq}. The model is built in a sequential manner with the same components as in SCNet.}
%\label{fig:1}
%\end{figure*}

\begin{figure*}[ht]
\centering
\includegraphics[width=1\linewidth]{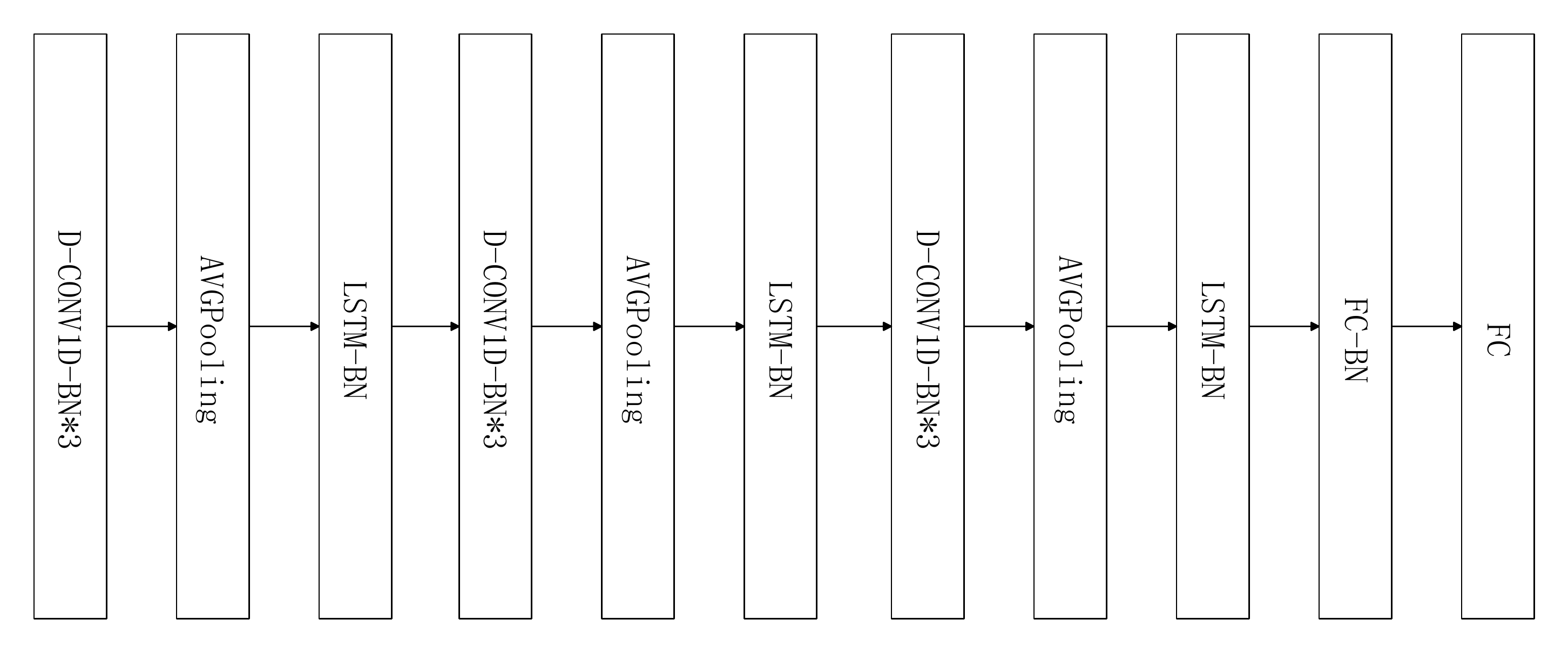}
\caption{The structure of SCNet\_{seq}. The model is built in a sequential manner with the same components as in SCNet.}
\label{fig:1}
\end{figure*}

\section{{SCNet}}
%add some attack evidence based on latent variable

%Our model uses the distribution of sampling points to predict the corresponding key byte. We use a train-to-learn strategy without presupposing a prior distribution. The model will learn it through the training data. 
To obtain the crossing information automatically, we design an encoder group and stack those having different hyperparameters to build an encoding block, which is powerful enough to extract the crossing information hidden in the traces. And the model has the capability to project the sampling points to multi-scale embedding information through the encoders. The embedding vectors between different sampling points in same trace usually are various, because at different times the chip may perform a different operation or just wait for the next operation. Moreover, the waveforms of different traces vary with the plaintexts and keys. When we fix the plaintext and key, the waveform is only related to noise corresponding to the chip. So the distribution of them is $P_{trace}(X|plaintext,key,noise)$. In addition, the distribution of sampling points in the traces is $P_{point}(x_{i}|X,i)$. And our model estimates a distribution:
\begin{align*} 
    P(key_{byte}|X)=P(labels|x_0,x_1,x_2,\dots,x_{M-1}),
\end{align*}
where $X$ is a trace consisted of sampling points $x_0,x_1,x_2,\dots,x_{M-1}$, and $key_{byte}$ is part of the key (one byte) used to produce these points. %Notice that the plaintext, key and noise are latent variables, but the plaintext and the key follow a uniform distribution. So the model needs to estimate a conditional probability distribution. 
To train the model, we adopt the multi-class cross-entropy function as the loss function:
\begin{equation*}
    Loss=-\frac{1}{BatchSize}\sum^{BatchSize-1}_{v=0}{\sum^{255}_{z=0}{\left(y_{v,z}{\mathrm{log} \left(y'_{v,z}\right)\ }\right)}}
\end{equation*}
since there are 256 ($2^8$) labels for each key byte. 

{In practice, one way to obtain the \textbf{\textit{crossing information}} is to utilize fully connected layers. Another way is to skip some points by using dilated encoder. Both methods have their advantages and disadvantages. The first method can completely obtain the information we need. However, too many parameters make the network prone to over-fitting on noise signal. The second method avoids using too many parameters. However, the dilated encoder may avoid too many crucial points, which makes it fail to predict the key. But in order to follow the conventional methods, we decided to adopt the dilated encoder. And using a group of dilated encoders together can make sure that all crucial points will make contribution to the final output.}

{As we mentioned, a group of dilated encoders is used to build a encoder block. For a dilated encoder, it can bypass some values and encode the rest. But it will be very unstable if we use a hyperparameter to decide which values to ignore according to their positions or features. It is much better to use convolution layers to calculate the crossing information, because the sliding window will obtain all combination situations if there are enough layers. To bypass some values, we can pad some zeros on filters.} We obtain the $i$-th output after performing an $r$-dilated encoding operation on the input $X$ with $l$-size kernel $f$:
\begin{equation*}
        {output}_i=\sum_{k=0}^{l-1}{x_{i+rk}*f_{k} },
\end{equation*}
%这个公式没问题吧
where $*$ is the convolution operator.

%\begin{table*}[ht]
%\centering
%\caption{A comparison between models. The models are trained on GTX TITAN X. "XXX Requires" is the minimum number of traces, which are required to get 100\%-correct forecast. "None" means that the model cannot get 100\%-correct forecast within 5000 traces.\ \protect ~\cite{standaert_unified_2009}}
%\resizebox{\textwidth}{!}{%
%\begin{tabular}{@{}rrrrrrr@{}}
%\toprule
%\textbf{Model Name} & \textbf{Training Time(s)} & \textbf{Model Size(MB)} & \textbf{ASCAD Desync0 Requires} & \textbf{ASCAD Desync50 Requires} & \textbf{ASCAD Desync100 Requires} & \textbf{DPA\_v4.2 Requires} \\ \midrule
%\textbf{ASCAD CNN} & 14250 & 508.0 & 150 & 4570 & None & None \\
%\textbf{SCNet\_seq} & \textbf{1000} & 26.5 & \textbf{80} & 1970 & \textbf{2760} & 1690 \\
%\textbf{SCNet} & 1675 & \textbf{16.6} & 160 & \textbf{530} & 3700 & \textbf{1200} \\ \bottomrule
%\end{tabular}%
%}
%\label{tab:1}
%\end{table*}

According to the property mentioned in Section~\ref{sec:2.2} of the Wiener process~\cite{stark_probability_2002}, we can use LSTM to denoise signal vectors. %In order to improve the stability of our model, we adopt the Inception and ResNeXt structure and increase the network cardinality~\cite{xie_aggregated_2017}. We also find that if we use a depth-wise separable convolutional layer~\cite{chollet_xception:_2017}, the network will be less stable. This is because that the bottleneck layer~\cite{lin_network_2013} adds noise as a result of parameter initialization in each channel. It will influence the LSTM layers. 
The structure of SCNet is shown in \textbf{Figure~\ref{fig:2}}. In SCNet, we incorporate six similar dilated encoder groups into each block to resolve sampling points at the same time. The LSTM layers reduce the number of feature maps after each block and ensure that the output channels are fewer than the feature maps of the next block to forget most of the duplicate and useless features.  %We reduce the width of the LSTM layers to forget most of the duplicate and useless features, and help the next block to further extract features. 

%In order to address these issues, we incorporate the Hybrid Dilated Convolution (HDC) technique~\cite{wang_understanding_2018} into SCNet to design dilation rates:
%\begin{equation*}
%    M_i=\max[M_{i+1}-2r_i,M_{i+1}-2\left(M_{i+1}-r_i\right),r_i],
%\end{equation*}
%where $r_i$ is the dilation rate of the $i$-th convolutional layer and $M_i$ is the maximum dilation rate of the $i$-th convolutional layer. There are ${n}$ layers in total and $M_n=r_n$. We increase the number of convolutional layer groups in each block in SCNet. We modify the dilation rate selection method for traces. In order to reduce the size of the filter, we need to increase the dilation rate to catch the sampling points which are distant from others. However, the biggest receptive field width cannot exceed $\frac{L}{4}$, where ${L}$ is the number of the sampling points. Thus, we need to fill the holes with ordinary convolutional layers in the last block in order to eliminate the gridding effect~\cite{wang_understanding_2018}. 

\begin{figure*}[h]
    \centering
    \begin{subfigure}[ht]{0.33\textwidth}
        \centering
         \includegraphics[width=1.0\textwidth,height=0.25\textheight]{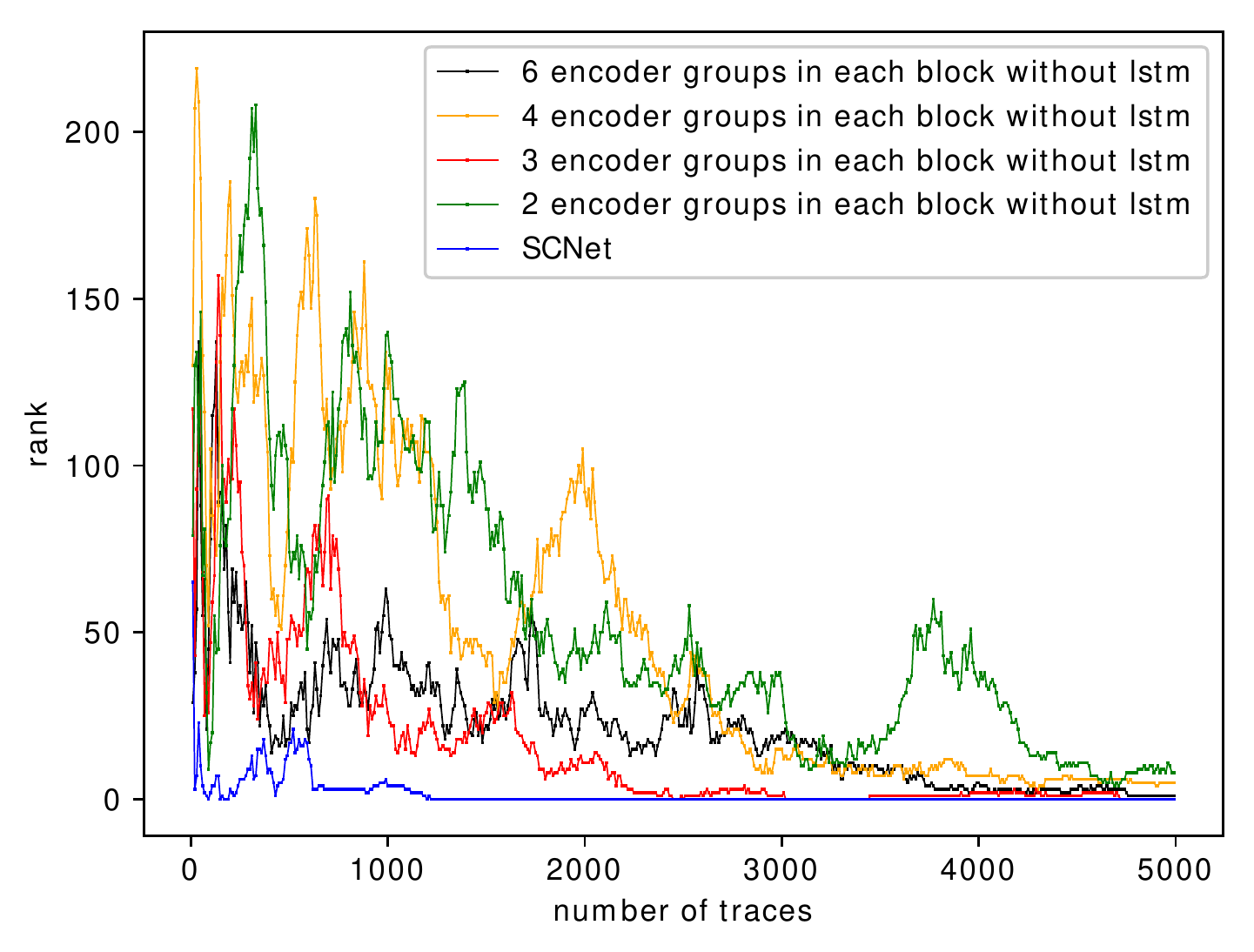}
         \caption{Different number of dilated encoder groups without LSTM.}
         \label{fig:abs_no_lstm}
    \end{subfigure}
    \begin{subfigure}[ht]{0.33\textwidth}
         \centering
        \includegraphics[width=1.0\textwidth,height=0.25\textheight]{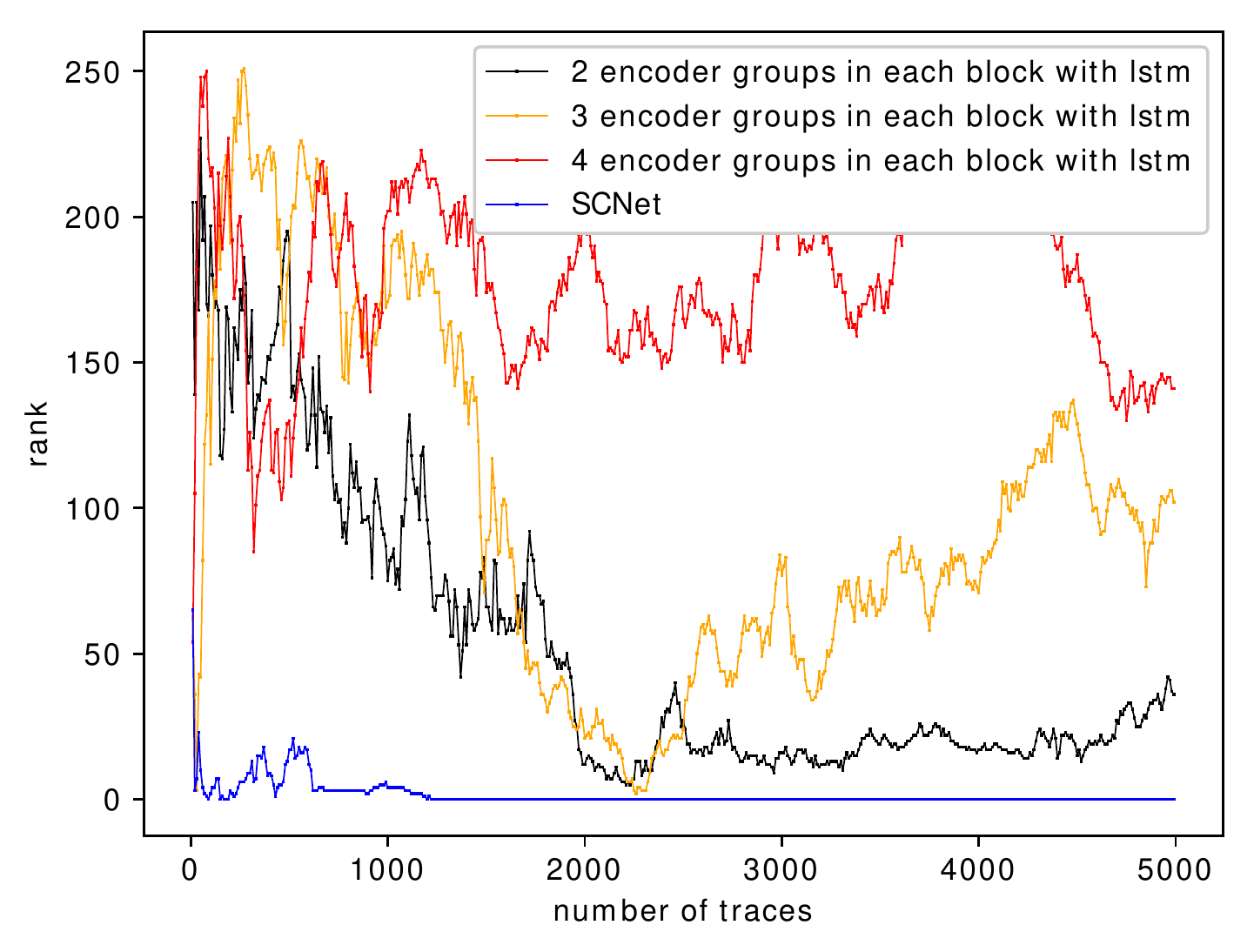}
         \caption{Different number of dilated encoder groups with three LSTMs.}
         \label{fig:abs_lstm}
    \end{subfigure}
    \begin{subfigure}[ht]{0.33\textwidth}
         \centering
         \includegraphics[width=1.0\textwidth,height=0.25\textheight]{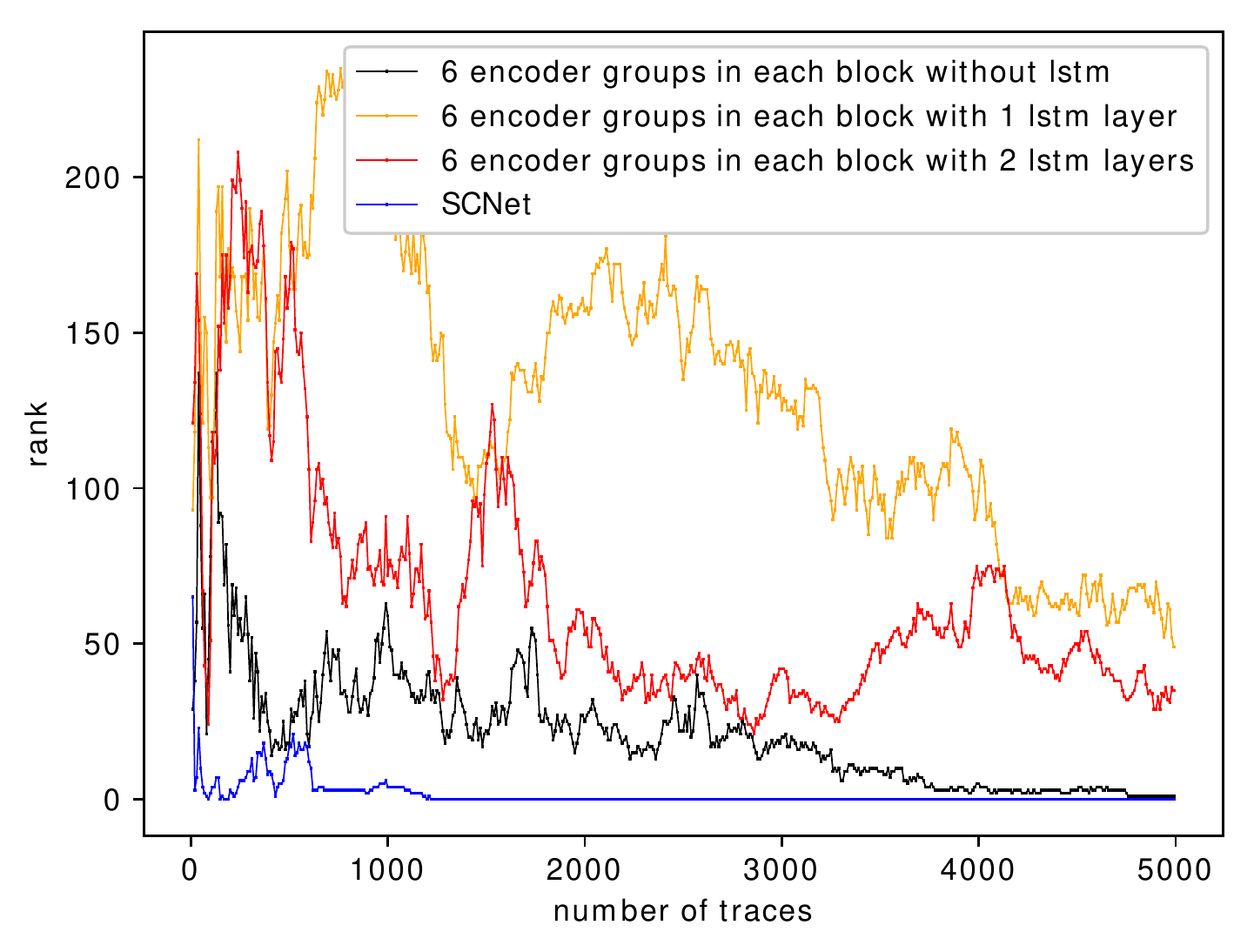}
         \caption{Different numbers of LSTMs with same encoder groups. }
         \label{fig:abs_different_lstm}
    \end{subfigure}
    \caption{Ablation study on DPA\_v4.2 dataset.}
\end{figure*}

\begin{table*}[ht]
\centering
\resizebox{380pt}{25pt}{
\begin{tabular}{@{}rrrrr@{}}
\toprule
\textbf{Model Name} & \textbf{Sliding Windows Size} & \textbf{Dilation Rate} & \textbf{LSTM Dimension} & \textbf{Feature Maps} \\ \midrule
\textbf{SCNet} & {[}7;7;7{]} & {[}15,13,11;9,7,5;5,2,1{]} & {[}16;64;256{]} & {[}8;32;92{]}\\
\textbf{SCNet\_seq} & {[}11;11;9{]} & {[}13,11,9;13,11,9;13,11,9{]} & {[}64;128;512{]} & {[}32,48,64;96,112,128;192,224,256{]}\\
 \bottomrule
\end{tabular}}
\caption{The Hyperparameters of SCNet\_seq and SCNet}
\label{tab:2}
\end{table*}

\begin{table*}[ht]
\centering
\begin{adjustbox}{max width=1.0\textwidth}
\begin{tabular}{@{}ccccccc@{}}
\toprule
                                  & \textbf{ASCAD CNN} & \textbf{SCNet\_seq} & \textbf{SCNet} & \textbf{6Group} & \textbf{3Group} & \textbf{Template Attack} \\ \midrule
\textbf{Training Time (sec)}      & 14,250             & 1,000               & 1,675          & 816             & \textbf{510}    & -                        \\
\textbf{Model Size (MB)}          & 508.0              & 26.5                & \textbf{16.6}  & 36.0            & 16.7            & -                        \\
\textbf{ASCAD Desync0 Required}   & 150                & \textbf{80}         & 160            & None            & None            & 190\cite{prouff_study_2018}                      \\
\textbf{ASCAD Desync50 Required}  & 4,570              & 1,970               & \textbf{530}   & None            & 4270            & 3200\cite{prouff_study_2018}                   \\
\textbf{ASCAD Desync100 Required} & None               & \textbf{2,760}      & 3,700          & None            & None            & None\cite{prouff_study_2018}                     \\
\textbf{DPA\_v4.2 Required}       & None               & 1,690               & 1,200          & 4760            & 3,200           & \textbf{10}              \\ \bottomrule
\end{tabular}
\end{adjustbox}
\caption{A comparison between models and traditional method template attack. The models are trained on GTX TITAN X. "\textbf{XXX Required}" is the minimum number of traces for dataset "\textbf{XXX}", which are required to get 100\%-correct forecast. "None" means that the model cannot get 100\%-correct forecast within 5000 traces.\ ~\protect\cite{standaert_unified_2009}}
\label{tab:1}
\end{table*}

\begin{figure*}[h]
    \centering
    \begin{subfigure}[ht]{0.33\textwidth}
        \centering
         \includegraphics[width=1.0\textwidth,height=0.25\textheight]{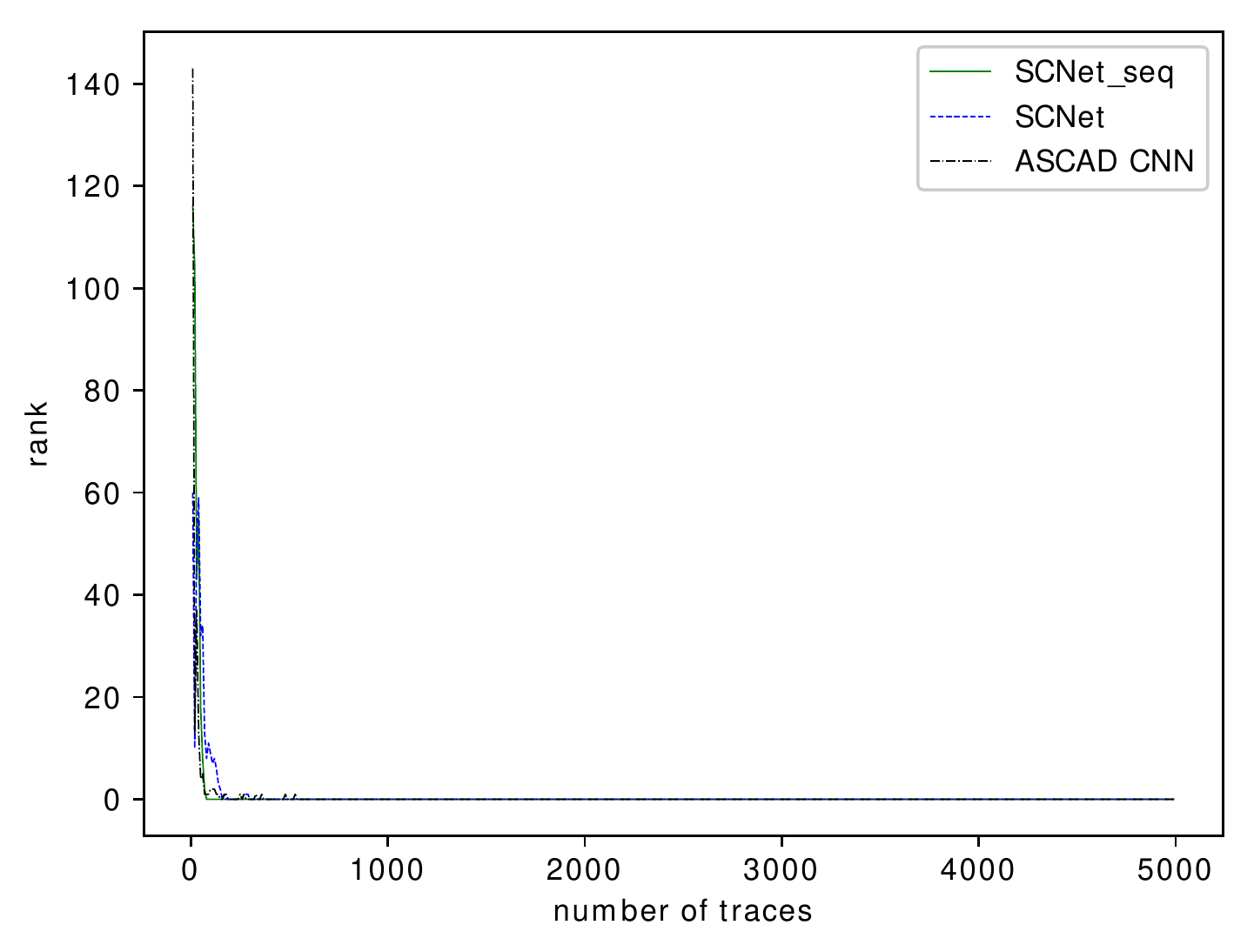}
        \caption{Different model's performance on the ASCAD Desync0 dataset.}
        \label{fig:3a}
    \end{subfigure}
    \begin{subfigure}[ht]{0.33\textwidth}
         \centering
        \includegraphics[width=1.0\textwidth,height=0.25\textheight]{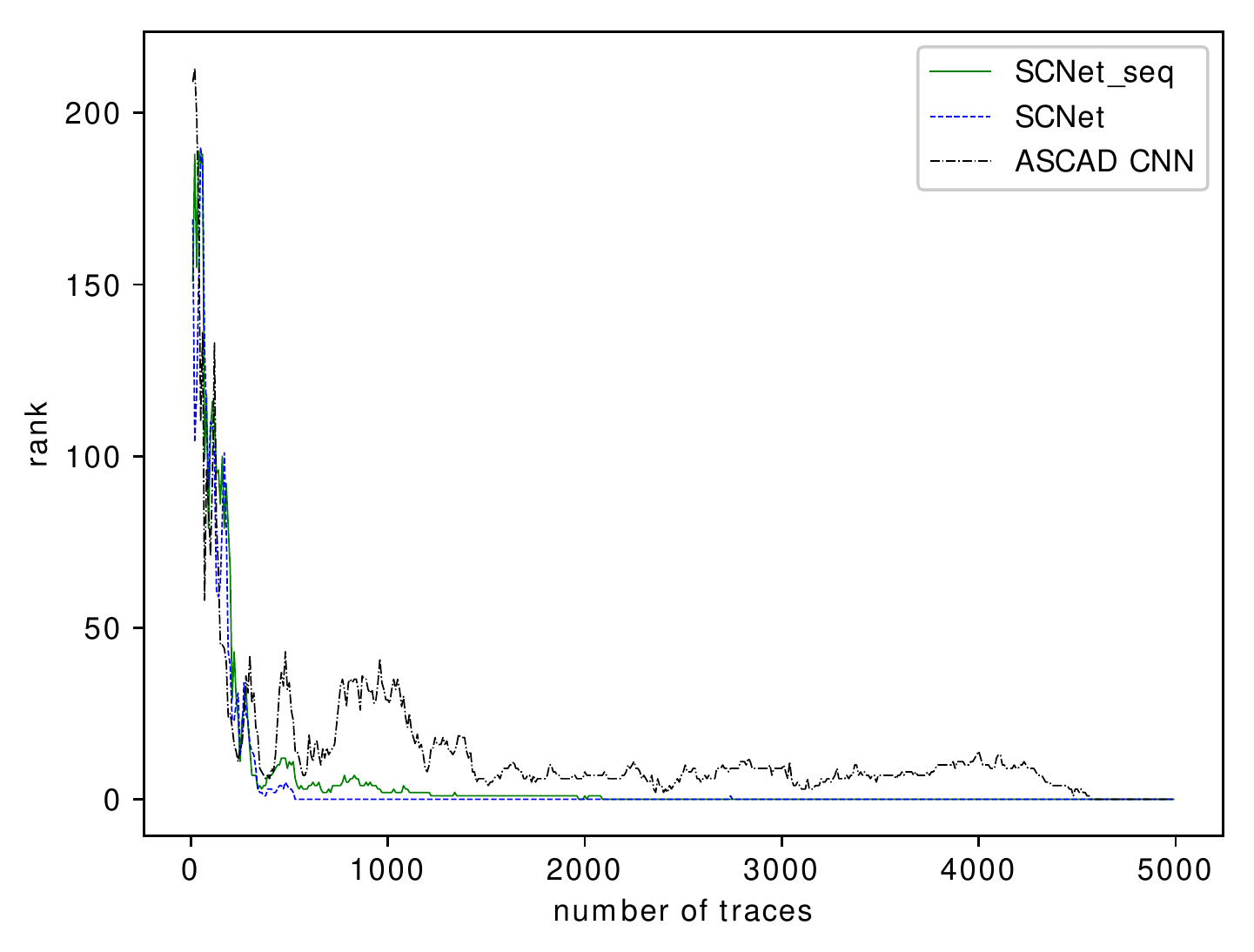}
        \caption{Different model's performance on the ASCAD Desync50 dataset.}
        \label{fig:3b}
    \end{subfigure}
    \begin{subfigure}[ht]{0.33\textwidth}
         \centering
         \includegraphics[width=1.0\textwidth,height=0.25\textheight]{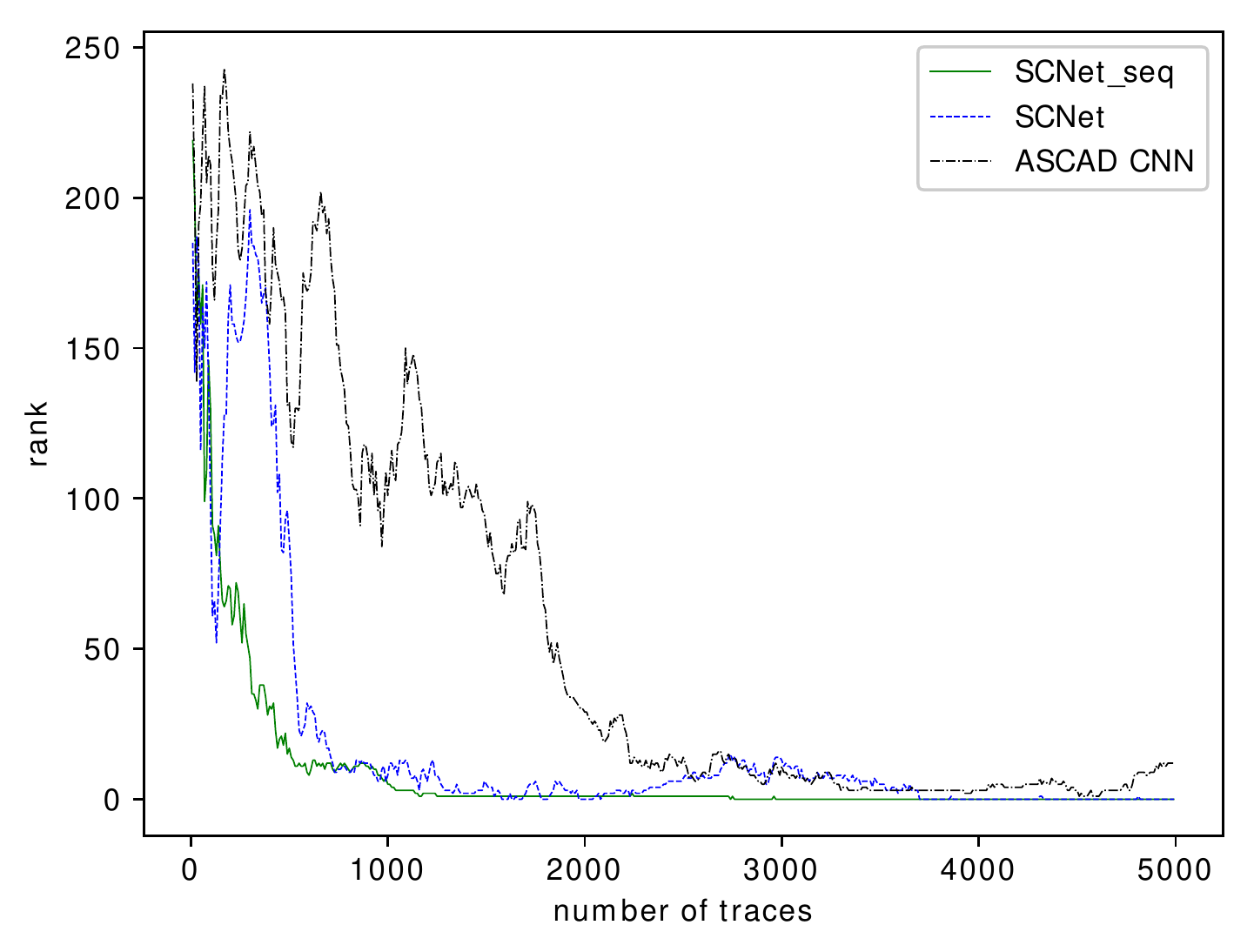}
        \caption{Different model's performance on the ASCAD Desync100 dataset. }
        \label{fig:3c}
    \end{subfigure}
    \caption{Different model's performance on the ASCAD dataset.}
\end{figure*}

\begin{figure}[ht]
\centering
\includegraphics[width=0.8\columnwidth,height=0.25\textheight]{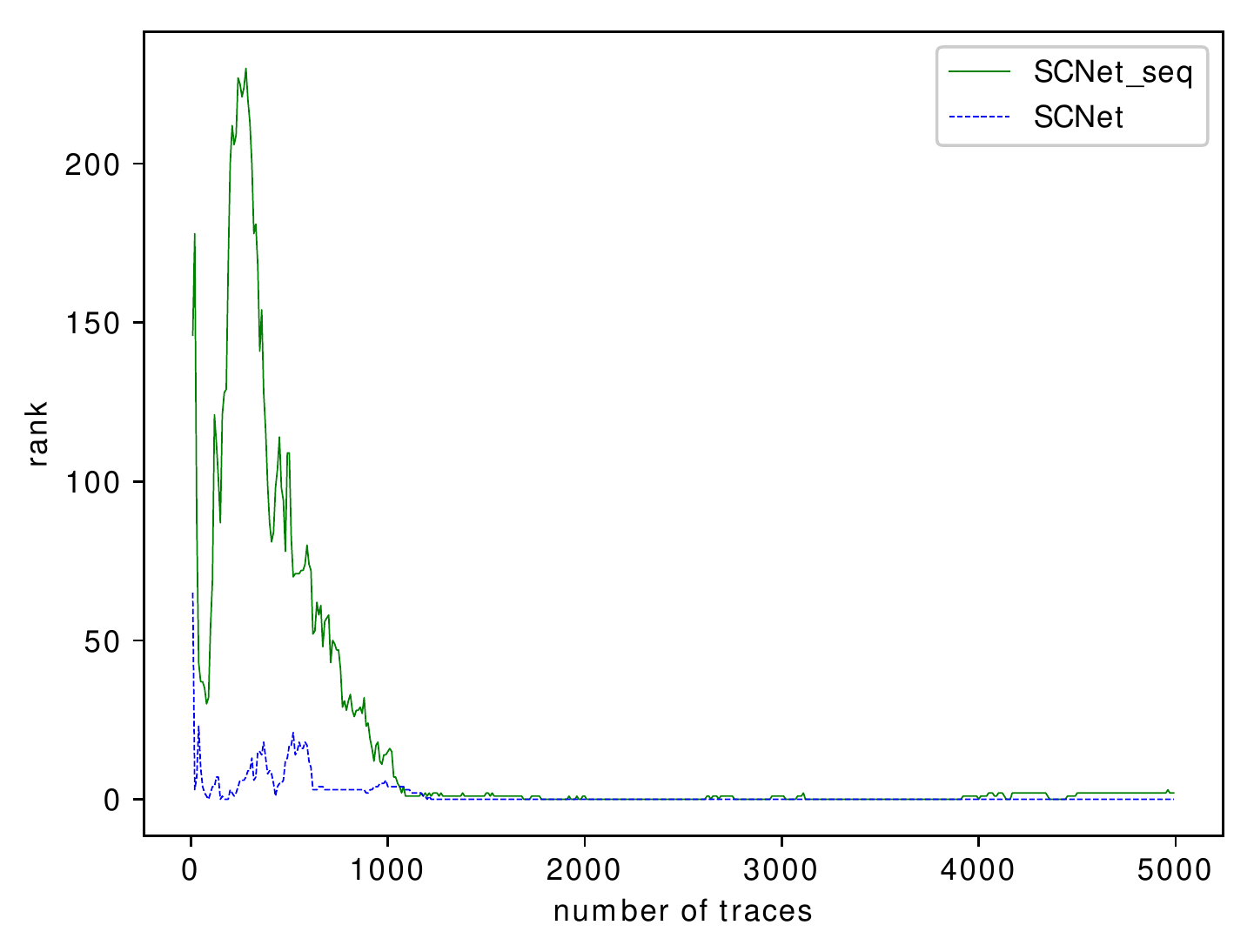}
\caption{The performance of different models on DPA\_v4.2.}
\label{fig:4}
\end{figure}

\section{Experimental Evaluation}
We conduct experiments based on real-world datasets to evaluate the performance of the proposed SCNet and compare with others.

\subsection{Datasets}
We use ASCAD dataset, which is a public side-channel dataset, to test the models. This dataset contains 50,000 items( i.e. traces) for training, which we split into the train set and the validation set; and another 10,000 items for testing. Each item has exactly 700 sampling points. All traces are generated by using different keys. The datasets has three versions with different offsets, which means these traces are not aligned and with most 0, 50 and 100 points deviation, respectively: 1) ASCAD Desync0, 2) ASCAD Desync50 and 3) ASCAD Desync100. It is suitable for evaluating the performance of models under complex conditions. 

We also use DPA\_v4.2~\cite{bhasin_analysis_2014} to test the models. We split the raw DPA\_v4.2 dataset into two parts. The first part contains 75,000 items for training, which we split into the train set and the validation set. The second part contains 5,000 items for testing. Each item has exactly 500 sampling points which are related to the 11-th byte of the AES secret key. In the train set, the items are generated by using 15 different keys. In the testing set, all the traces are generated by using a same key. Both of the datasets are acquired by the software implementing the AES algorithm with a mask.

%\begin{figure}[!ht]
%\centering
%\includegraphics[width=250pt,height=220pt]{abs1}
%\caption{Ablation study on DPA\_v4.2 dataset. Different number of dilated convolution groups without LSTM.}
%\label{fig:abs_no_lstm}
%\end{figure}

%\begin{figure}[!ht]
%\centering
%\includegraphics[width=250pt,height=220pt]{abs2}
%\caption{Ablation study on DPA\_v4.2 dataset. Different number of dilated convolution groups with three LSTMs.}
%\label{fig:abs_lstm}
%\end{figure}

%\begin{figure}[!ht]
%\centering
%\includegraphics[width=250pt,height=220pt]{abs3}
%\caption{Ablation study on DPA\_v4.2 dataset. Different numbers of LSTMs with same convolution groups. }
%\label{fig:abs_different_lstm}
%\end{figure}

\subsection{Comparison Models}
Since we evaluate our model on ASCAD as a main result, it is necessary to compare with a well designed and trained neural network as well as a traditional template attack. Thus, we adopt the ASCAD CNN as the comparison model, in order to illustrate the effects of our design choices. %It is a modified AlexNet with the 2-D convolutional layers replaced with the 1-D convolutional layers and using the ReLU activation function. 
Another baseline model is SCNet\_seq, which is showed in \textbf{Figure~\ref{fig:1}}. It employs all the components in SCNet, but the whole model is built in a sequential manner. It has less dilated encoders and wider LSTM layers than SCNet. Additionally, to test the effect of LSTM layers, we compare SCNet with \textbf{6Group} and \textbf{3Group}. 6Group has the same block structure as SCNet, but does not contain any LSTM layer. 3Group reduces the number of dilated encoder groups in each block by 50\%, and with no LSTM layer. The parameters of SCNet and SCNet\_seq are shown in \textbf{Table~\ref{tab:2}}.

\subsection{Ablation Study}
To validate the effectiveness of SCNet, we perform ablation experiments on DPA\_v4.2 dataset. In all experiments, we use the same optimization hyperparameters and split the train data into the train set and the validation set with same random seed. In order to limit the size of model and computational complexity as well as prevent over-fitting, we use 3 encoder blocks in the network at most. In each block, there are 6 dilated encoder groups at most. In each group, there are three convolution layers.

%\begin{figure}[ht]
%\centering
%\includegraphics[width=1\columnwidth]{abs1}
%\caption{Ablation study of different number of dilated convolution groups without using LSTM.}
%\label{fig:abs_no_lstm}
%\end{figure}

%\begin{figure}[ht]
%\centering
%\includegraphics[width=1\columnwidth]{abs2}
%\caption{Ablation study of different number of dilated convolution groups using three LSTM layers.}
%\label{fig:abs_lstm}
%\end{figure}

%\begin{figure}[ht]
%\centering
%\includegraphics[width=1\columnwidth]{abs3}
%\caption{Ablation study of six dilated convolution groups using different numbers of LSTM layers.}
%\label{fig:abs_different_lstm}
%\end{figure}

More specifically, we evaluate networks using the same sliding windows size and dilation rate without the LSTM layers at first. The experiments study the effect of different number of dilated encoder groups in each block, as shown in \textbf{Figure~\ref{fig:abs_no_lstm}}. ``rank'' is the position where the correct byte appears in the output, which is sorted in descending order of the probability~\cite{standaert_unified_2009}. The smallest number of traces that can reach rank 0 is the only metric in our experiments. The best result of the network without using LSTM layers is achieved by the network with three dilated encoder groups. {Even we use same number of encoder groups, the one without LSTM layers can not beat the SCNet. And it is clear that without LSTM layers, the prediction curves do not decline as smoothly as the one of SCNet. This means that it is hard for the models to decrease the guess entropy by using more traces, directly. The LSTM layers can make sure that the model predicts unchanging distribution so that when predicting on more traces, it achieves better results.} %More dilated convolution groups produce worse results. We can use LSTM layers to achieve better performance. 

Then, we evaluate networks with three LSTM layers. The experiments study the effect of the number of dilated encoder groups in each block, as shown in \textbf{Figure~\ref{fig:abs_lstm}}. The results achieved by the networks with between 2 to 4 groups are not good. SCNet, which uses 6 groups in each block, achieves the best result. {For networks with less groups in each block, the crossing information hidden behind the traces is hard to exact. And the lowest rank appears in each predict curve is related with the number of groups. The more groups are used, the less traces are needed to achieve the lowest point. However the lowest point is not related with the number of groups. And the model needs more groups to obtain the crossing information we need to predict the distribution.}%It captures the special properties of the side-channel data better than other approaches. 

Finally, we evaluate networks which have the same number of dilated encoder groups in each block. The experiments study the effect of different numbers of LSTM layers. The results are shown in \textbf{Figure~\ref{fig:abs_different_lstm}}. %When we use one or two LSTM layers after the first block or the first and the second block, it will influence the feature maps in the last block, and will produce worse results. 
{For networks with less LSTM layers, it is hard to denoise the traces. And the lowest rank appears in each predict curve is related with the number of LSTM layers, which is quite similar with the relation of the number of groups. The more LSTM layers are used, the less traces are needed to achieve the lowest point. However the model without any LSTM layer has better result than the one with one or two LSTM layers. When the model is equipped with not enough LSTM layers, the quailty of the prediction distribution is not as good as the one without LSTM layers. And the model needs as much as possible LSTM layers to guarantee that the noise is denoised.}

\subsection{Results}
We perform extensive experiments comparing the ASCAD CNN, template attack and our proposed models. \textbf{Table~\ref{tab:1}} shows a detailed comparison between them. SCNet\_seq and SCNet have fewer parameters than ASCAD CNN (\textbf{26.5MB} and \textbf{16.6MB} \textit{vs} \textbf{508.0MB} with an almost \textbf{95\%} reduction)
and are faster in terms of training (\textbf{1,000s} and \textbf{1,675s} \textit{vs} \textbf{14,250s} with an almost \textbf{90\%} reduction).
To compare the results, we use the same test code which is provided by~\cite{prouff_study_2018}. Both of two models achieve significantly better performance than ASCAD CNN even template attack. For the ASCAD Desync0 dataset, which is an aligned dataset, our two networks only need to use around a hundred traces(\textbf{80} or \textbf{160}) 
to obtain the key byte. More traces(\textbf{1,970} or \textbf{530}) 
are needed for the ASCAD Desync50 dataset, which is a lightly unaligned dataset. Even more traces(\textbf{2,760} or \textbf{3,700}) 
are needed for the ASCAD Desync100 dataset, which is a heavily unaligned dataset. For the DPA{\_v4.2} dataset, which is an aligned dataset, around one thousand traces are needed(\textbf{1,690} or \textbf{1,200}).
%The results are shown in \textbf{Figure~\ref{fig:3a}}, \textbf{Figure~\ref{fig:3b}}, \textbf{Figure~\ref{fig:3c}} and \textbf{Figure~\ref{fig:4}}. 

{To increase the credibility of the results, we also compare SCNet with 6Group and 3Group. And the results shows that the SCNet has better performance on all datasets. The 6Group and the 3Group can only achieve relatively good results on DPA\_v4.2. And on ASCAD, all three sub-datasets are extremely challenging for the 6Group and the 3Group. }

%\begin{figure}[ht]
%\centering
%\includegraphics[width=1\columnwidth]{image3}
%\caption{The performance of ASCAD CNN, SCNet\_seq and SCNet using the ASCAD Desync0 dataset.}
%\label{fig:3a}
%\end{figure}

%\begin{figure}[ht]
%\centering
%\includegraphics[width=1\columnwidth]{image4}
%\caption{The performance of ASCAD CNN, SCNet\_seq and SCNet using the ASCAD Desync50 dataset. }
%\label{fig:3b}
%\end{figure}

%\begin{figure}[ht]
%\centering
%\includegraphics[width=1\columnwidth]{image5}
%\caption{The performance of ASCAD CNN, SCNet\_seq and SCNet using the ASCAD Desync100 dataset. }
%\label{fig:3c}
%\end{figure}

{\textbf{Figures~\ref{fig:3a},~\ref{fig:3b},~\ref{fig:3c}} show the performance of SCNet\_seq, SCNet and ASCAD CNN on the three ASCAD datasets, respectively. In \textbf{Figure~\ref{fig:3a}}, there is no significant difference between the number of traces the three models use to predict the key correctly. All curves are very smooth without large fluctuations. However, the biggest value of guessing entropy achieved by SCNet is lower than those of the other two approaches. In \textbf{Figure~\ref{fig:3b}}, SCNet achieves significantly better results than the others. And the curve of ASCAD CNN repeatedly rises and falls sharply between 300 traces and 1,500 traces. Between 1500 traces and 4,000 traces, it has a small change. In \textbf{Figure~\ref{fig:3c}}, SCNet\_seq is better than SCNet and ASCAD CNN. The ASCAD CNN is the worst one, which can not reduce the guessing entropy to 0 in 5,000 traces. Although the ASCAD CNN curve declines, it remains within a relatively stable range finally. And the one of SCNet falls quickly at first, but it rises to a very high value and falls again. After fluctuating within a certain range, it reduces to 0 when using 3,700 traces and stays at 0. The SCNet\_seq outperforms others on this dataset. The curves of it falls quickly to 0 after using 2760 traces which is about 1,000 traces less than SCNet and stays at 0 stably. In \textbf{Figure~\ref{fig:4}}, we only show results of SCNet\_{seq} and SCNet, while the ASCAD CNN cannot provide the correct prediction after training. The SCNet achieves better result than SCNet\_seq. The curve of SCNet shows that the guessing entropy at start is very low and it falls to 0 quickly after using 1,200 traces and does not rise again. But the SCNet\_seq one is very unstable at first. It starts at a higher point and falls firstly but then rises up very fast. After rising to the highest point, it falls to 0 slowly. Finally, it stays at 0 after using 1,690 traces. Additionally, it is much easier to train models on datasets with smaller offsets, as expected.}
%SCNet achieves lower GE and higher SR than SCNet\_seq.  On datasets with large offsets, it is difficult to ensure that models can predict the results correctly. If the parameters are not initialized well, it may result in the gradient vanishing problem, making it harder to train the network.

\section{Conclusions and Future Work}
In this paper, we propose SCNet. It takes the sampling point sequences as the input and obtains the key byte, which is used in the block cipher algorithms. To obtain the key byte, SCNet adopts dilated encoders and sampling point embedding vectors to capture the cross information (high order relationship). Extensive experiments on the ASCAD and the DPA\_v4.2 datasets show that SCNet can restore key bytes with significantly fewer traces than ASCAD CNN. The outcomes from this research is very useful for both attack and defense in block cipher security.

And in the future, our model may could be used to obtain the data running in GPUs when people train other models. This is a huge threat for those sensitive information and privacy. In future research, we will study how to adapt SCNet to compromise the security of today's federated machine learning (FML) systems, thereby finding ways to improve their robustness.

%As the future of AI is moving towards increasingly emphasizing preserving user privacy and data confidentiality, federated machine learning (FML) has been proposed as a new paradigm of machine learning and is being adopted more widely as time goes by. In FML, data owners train local models based on their local datasets and share the model parameters with a centralized entity %called the data federation following some privacy-preserving secure protocols 
%in order to train a more effective collective machine learning model without really sharing private data. In future research, we will study how to adapt SCNet to compromise the security of today's FML systems, thereby finding ways to improve their robustness.

%\begin{acks}
%This research is supported, in part, by Nanyang Technological University, Nanyang Assistant Professorship (NAP).
%\end{acks}

\bibliographystyle{ACM-Reference-Format}
\balance
\bibliography{bib}

\end{document}